\documentclass[12pt,preprint]{aastex}
\usepackage{lscape}
\def\kms{{\rm km\,s^{-1}}}

\def\masyr{{{{\rm mas}}\,{{\rm yr}}^{-1}}}
\def\usno{{\rm USNO}}
\def\rnltt{{\rm rNLTT}}
\def\nltt{{\rm NLTT}}
\def\sdss{{\rm SDSS}}

\def\lim{{\rm lim}}

\begin{document}

\title{Halo Luminosity Function From
Photometric Calibration of the Revised NLTT}

\author{Andrew Gould, Juna A.\ Kollmeier, Julio Chanam\'e}
\affil{Department of Astronomy, The Ohio State University,
140 W.\ 18th Ave., Columbus, OH 43210}
\authoremail
{gould,jak,jchaname@astronomy.ohio-state.edu}
\author{Samir Salim}
\affil{Department of Physics and Astronomy,
University of California at Los Angeles,
Los Angeles, CA 90095}
\authoremail
{samir@astro.ucla.edu}

\singlespace

\begin{abstract}

We calibrate the photographic photometry
of the revised New Luyten Two-Tenths catalog (rNLTT)
by matching 3448 rNLTT stars to the Sloan Digital Sky Survey (SDSS).
The correction is linear in magnitude and goes from zero at $V\sim 14$
to 0.32 mag at $V=19$, in the sense that rNLTT was previously too bright.
The correction implies that the underlying USNO-A2.0 photometry, on
which rNLTT photometry is based, is non-linear.  The new calibration
somewhat improves the appearance of the $(V,V-J)$ reduced proper
motion diagram in the sense of better separation between disk and
halo stars.  We repeat Gould's analysis of 5000 halo stars in rNLTT.
The most important change is to move the peak of the halo luminosity
function about 0.5 mag dimmer, from $M_V=10.5$ to $M_V=11$, putting
it into good agreement with the parallax-based determination of
Dahn et al.

\end{abstract}
 
\section{Introduction
\label{sec:intro}}

	The revised NLTT (rNLTT), assembled by \citet{gould03bright}
and \citet{salim03}, presented improved astrometry and photometry for 
approximately 36,000 stars drawn from the New Luyten Two Tenths 
catalog (NLTT)
\citep{luyten7980}.  At the bright end, NLTT stars were matched primarily 
to the Hipparcos \citep{esa97}
and Tycho-2 \citep{t2} catalogs, while at the faint end they were matched
to USNO-A \citep{usnoa2} and 2MASS \citep{2mass}.  
The bright end covers the whole sky. The faint
end basically covers the 44\% of the sky represented by the intersection 
of the 2MASS Second Incremental Release
and First Palomar Observatory Sky Survey (POSS I), as reduced and cataloged 
by USNO-A.
The roughly 45 year baseline between these observations
enabled a proper-motion precision of $5.5\,\masyr$, roughly a 4-fold
improvement over NLTT.  By identifying USNO-A and 2MASS counterparts, rNLTT
assigned $(V,J)$ magnitudes to almost all entries, with $V$ being derived
from USNO-A photographic photometry in $B_\usno$ and $R_\usno$ and 
with $J$ coming
directly from 2MASS.  This was a huge improvement over the original
$(B_\nltt,R_\nltt)$ photographic photometry of NLTT, in part because USNO-A
photographic photometry has smaller scatter than Luyten's, but primarily
because $V-J$ provides a much longer color baseline than does
$B_\nltt-R_\nltt$.  Indeed, using a $(V,V-J)$ reduced proper motion (RPM)
diagram, \citet{salim02} were able for the first time to cleanly
separate main-sequence, subdwarf, and white dwarf stars in NLTT.  This
clean separation then permitted \citet{gould03halo} to analyze a sample
of 4500 halo stars and to measure their luminosity function and
velocity distribution with much greater precision than was previously
possible.  The improved astrometry and photometry of rNLTT also
permitted \citet{chaname04} to assemble a catalog with more than 1000
wide binaries, also cleanly separated into disk and halo objects.

	The photometric calibration of the rNLTT optical $(V)$ band
was derived by comparing USNO-A photometry of NLTT stars with  
photometry found in the literature for the same stars, including
36 white dwarfs from \citet{ldm88} and 19 M dwarfs from \citet{weis96}.  
While the sample covered a broad range of magnitudes, $11\la V \la 18$,
the small sample size led \citet{salim03} to solve only for a
zero point and a color term but to assume that the flux levels
derived from USNO-A magnitudes were linear in flux.  A proper calibration
would require a sample that is both large and covers the full
range $12\la V \la 19$ from the magnitude limit of Tycho-2 
(and hence the onset of USNO-A-based photometry)
to the magnitude limit of rNLTT.

	The Sloan Digital Sky Survey (SDSS) 
Data Release 3\footnote{http://www.sdss.org/dr3}
\citep{abaz05}
now provides such a sample
and so permits a calibration of rNLTT photometry over the entire range
$12\la V \la 19$.  We carry out this calibration and find that rNLTT
assigned $V$ magnitudes that are too bright by an amount that 
grows linearly with increasing $V$.  That is, $\Delta V = 0.0633(V-13.9)$.
Because this offset is of order the statistical photometric error of
0.22 mag, it is too small to have any practical impact on applications
that use rNLTT to locate interesting classes of stars, such as 
subdwarfs \citep{yong03}, white dwarfs \citep{vennes03,kawka04},
red dwarfs \citep{reid04}, and dwarf carbon stars \citep{lowrance03}.
However, because the effect is systematic, it could potentially affect
studies that derive parameters from a statistical analysis of rNLTT.
In this case, the fractional errors in the derived parameters can be
much smaller than the errors associated with individual stars.

	In some respects, rNLTT has already been superseded by
the \citet{lepine05} proper motion catalog (LSPM), which covers
the entire northern sky, $\delta\geq 0$.  LSPM has a proper-motion
threshold of $\mu\geq 150\,\masyr$ (compared to $180\,\masyr$ for NLTT)
and is more complete than NLTT (and indeed is virtually 100\% complete
to $V = 19$).  Ipso facto, LSPM is more complete than rNLTT in
the region it covers.  However, rNLTT has significantly smaller
astrometric errors (5.5 vs.\ 8 $\masyr$) and somewhat smaller photometric
errors and also covers areas south of the equator.  While all three of these
shortcomings can be rectified in principle,
substantial additional work will be required.
Moreover, the photometry and astrometry improvements would be achieved
by cross-identifying LSPM stars with USNO-A, and when this is done,
the photometric calibration presented here will still have to be applied.

	Finally, \cite{gould03halo}'s statistical analysis of rNLTT halo stars
will form a baseline of comparison for future studies based on even
larger proper-motion samples.  It is therefore important to remove
systematic effects in this analysis.

\section{Calibration
\label{sec:calib}}

	Our basic plan of attack is to cross identify rNLTT with SDSS
(third data release) in order
to plot $\Delta V \equiv V_\sdss - V_\rnltt$, where 
\begin{equation}
V_\sdss \equiv r + 0.44(g - r) - 0.02
\label{eqn:vsdss}
\end{equation}
is the transformation of SDSS magnitudes into Johnson $V$ as s
calibrated by \citet{juric02} and
\begin{equation}
V_\rnltt \equiv R_\usno + 0.32(B-R)_\usno + 0.23
\label{eqn:vrnltt}
\end{equation}
is the transformation from USNO-A2 magnitudes into Johnson V as used in rNLTT.
\citet{juric02} estimate that the SDSS calibration reproduces Johnson $V$
to ``better than $\sim 0.05$ mag''.
Before proceeding, we note that equation (\ref{eqn:vrnltt}) applies directly 
only for stars $\delta>-20^\circ$, where rNLTT derives its photometry
and astrometry from USNO-A2.0.  For $-33^\circ\la\delta \la -20^\circ$,
rNLTT uses USNO-A1.0, which requires a photometric conversion given
by \citet{salim03} before applying equation (\ref{eqn:vrnltt}).
However, since all SDSS areas lie in the former region, this photometric
conversion plays no role in the current work.

	In order to assure a clean sample, we restrict the selection
as follows.  First, we consider only SDSS stars with proper-motion entries, 
derived from USNO-B1.0 \citep{monet03} cross identifications,
$\mu>140\,\masyr$.  Second, we exclude SDSS stars with a ``bad''
position on the $gri$ color-color diagram.  Here ``bad'' means away
from the characteristic ``dog-leg'' stellar locus of this diagram,
except that we allow stars with degenerate M-dwarf $g-r$ colors
($1.2<g-r<1.8$) but with anomalously low $i$ flux due to saturation.
Third, we consider only rNLTT stars
that are matched to both USNO-A and 2MASS and with $V_\rnltt>12$.
Fourth, we exclude rNLTT
stars with binary companions within $10''$.
Fifth, we demand that the rNLTT epoch 2000
position agree with the SDSS position to within $3''$.  Finally, we
exclude matches with discrepant proper motions (vector difference greater 
than $30\,\masyr$), unless their positions lie within $1''$.

	SDSS (by virtue of USNO-B) and rNLTT (based on USNO-A) both use
POSS I for the first epoch of their
proper-motion estimates, so the first condition implies a high
probability that if there is a position match, it is to a genuine
rNLTT star.  SDSS photometry can be compromised by saturation, and
this is particularly a worry for stars at the bright end of our 
investigation.  The $gri$ test should exclude most saturated
stars.  However, if the $g-r$ color is consistent with the degenerate
M dwarf track, we do not want to exclude the star simply because
the $i$ measurement is corrupted.  This is to be expected for
bright, red stars, but does not interfere with our $V$ band
estimate (which depends only on $g$ and $r$).  The third condition
is motivated by the fact that while rNLTT does
contain entries that are missing either USNO-A or 2MASS data,
these identifications are generally less secure. Moreover, in the
former case, of course, its $V$ estimate is not based on USNO-A
photometry.  The fourth condition is imposed because USNO-A
photometry can be corrupted by the presence of a near neighbor.
In general, we allow for a $3''$ mismatch in position because
SDSS positions are given at the epoch of observation while
rNLTT positions are at epoch 2000.  For high proper-motion stars
there can be a significant offset.  However, we guard against
random matches by demanding $1''$ agreement unless the proper motions
differ by less than $30\,\masyr$.  

\begin{figure}
\plotone{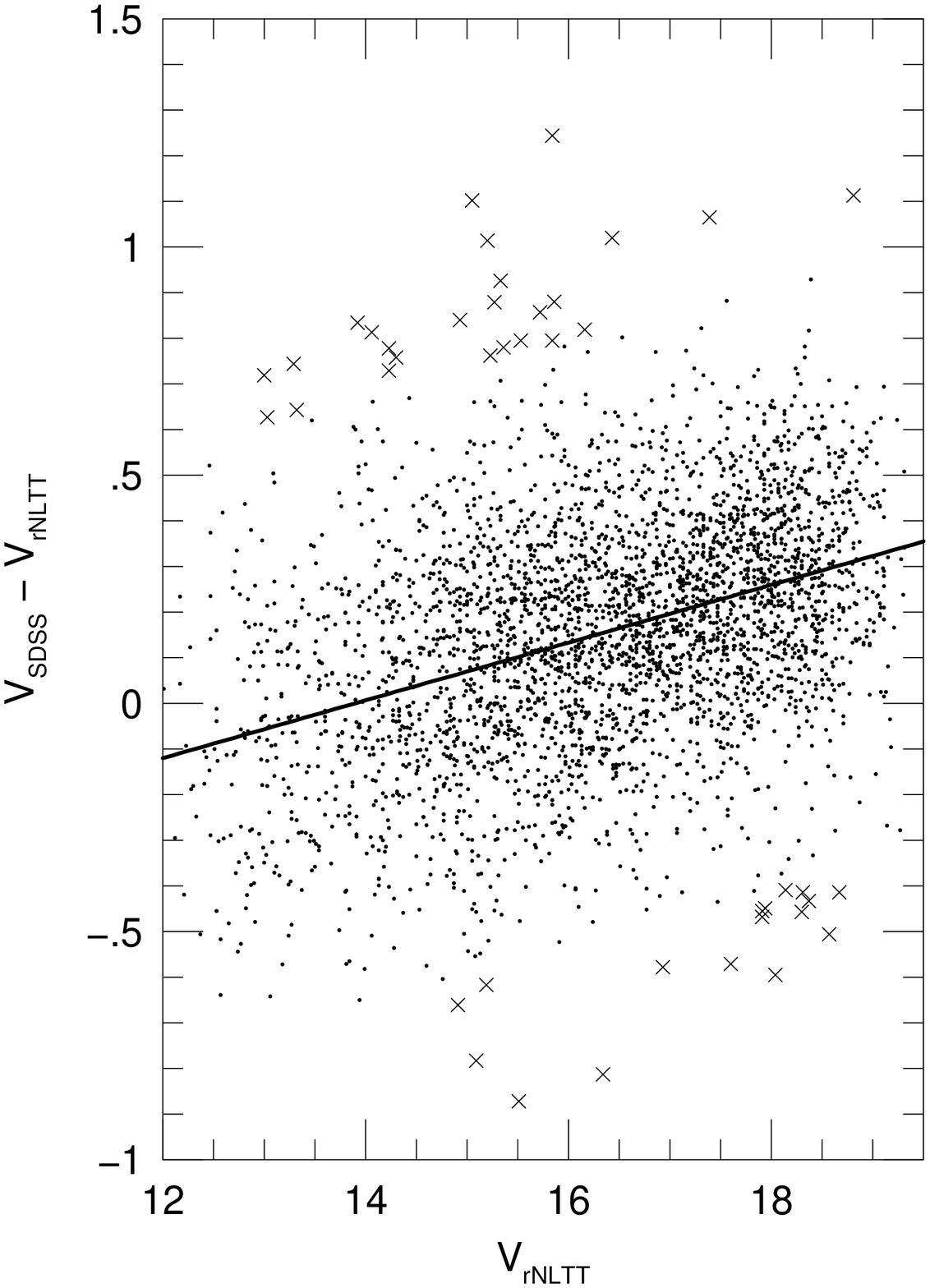}
\caption{\label{fig:calib}
Difference ($\Delta V = V_{\rm SDSS} - V_{\rm rNLTT}$) between
$V$ band measurements as derived from SDSS CCD photometry and
rNLTT (ultimately USNO-A) photographic photometry for 3448
stars in common. Crosses indicate the 51 recursively removed $3\,\sigma$ 
outliers. The remaining points are fit to
a straight line $\Delta V = 0.1331 + 0.0633(V_\rnltt - 16)$,
which is shown in bold.  The residual scatter of these 3397 points
is 0.22 mag.
}\end{figure}

	In Figure \ref{fig:calib}, we plot 
$V_\sdss - V_\rnltt$ for the sample of 3448 matches obtained in this
way.  We fit these points to a straight line and estimate the
error from the scatter by demanding that $\chi^2$ per degree of freedom 
(dof) be equal to unity.  We eliminate the largest outlier and repeat this
procedure until the largest outlier is less than $3\,\sigma$.
This results in 51 eliminations compared to 10 that would be expected
from a strictly Gaussian distribution.  That is, non-Gaussian outliers
constitute 1\% of the sample.  We find,
\begin{equation}
V_\sdss - V_\rnltt = (0.1331 \pm 0.0039) + (0.0633\pm 0.0024)(V_\rnltt -16),
\label{eqn:Vcor}
\end{equation}
with a scatter of $\sigma=0.22\,$mag,
where the zero-point offset $(V_\rnltt -16)$ is chosen to minimize 
(among integer values) the correlation between the two terms.
This relation implies that relatively faint stars $(V>14)$ should
be corrected to still fainter $V$ magnitudes, and hence also redder 
$V-J$ colors. Because the slopes of main sequence and subdwarf sequence
are greater than unity, the color change has a bigger impact, so that
both of these sequences become {\it brighter} at fixed color.

	We try extending this linear fit to next order, but find that 
the additional quadratic term does not significantly improve the fit.
Similarly, we find no correlation between the residuals to the fit
and observed $V-J$ color.

	Equation (\ref{eqn:Vcor}) implies that USNO-A2.0 photometry
is non-linear.  By combining this equation with equation (\ref{eqn:vrnltt}),
we derive an explicit expression for this nonlinearity,
\begin{equation}
V_\sdss = 1.0633\,R_\usno + 0.3403(B-R)_\usno -0.6351
\label{eqn:vusno}
\end{equation}

	While SDSS photometry is quite homogeneous, this is not
necessarily true of USNO-A photometry, which is derived from photographic
plates exposed under strongly varying conditions.  Our principal
concern is not plate-to-plate variations, which would be too localized
to affect most statistical applications, but possible broad trends
with position on the sky.  We therefore divide the data into seven
different subsets, the four quadrants in right ascension and three
bands in declination (separated at $\delta=0^\circ$ and $\delta=40^\circ$).
The only significant differences that we find are for the $(\delta<0^\circ)$
and $(0^\circ<\delta<40^\circ)$ subsamples, which have zero points that
are respectively about 0.035 mag greater and 0.025 mag smaller than the
sample as a whole.  While these deviations are detected with high
(almost $7\,\sigma$) confidence, they are quite small compared to the
scatter of 0.22 mag.  We will ignore them in what follows, but for
some applications the reader should be aware that they are present.

Figure \ref{fig:rpm} shows the rNLTT $(V,V-J)$ RPM diagram before 
and after correcting the photometry.  
What is plotted is not the traditional RPM but
the parameter
\begin{equation}
V_{\rm rpm,adj} \equiv V + 5\log\mu - 1.47|\sin b| - 2.73,
\label{eqn:vrpmdef}
\end{equation}
where $\mu$ is the proper motion in arcsec per year and $b$ is 
the Galactic latitude.  As shown by \citet{salim03}, adding the
Galactic-latitude term yields a cleaner separation of disk and halo
stars.  The line, $\eta=0$, where
\begin{equation}
\eta \equiv V_{\rm rpm,adj} - 3.1(V-J)
\label{eqn:etadef}
\end{equation}
is the boundary separating these 
populations that was adopted by \citet{salim03}.  The principal
change is that stars in the lower portions of the diagram are
shifted to the red because the $V$ measurements are now fainter.
(Of course, they are also shifted down, but this is less noticeable.)\ \
In addition, the ``trough'' between the disk and halo tracks now
appears both straighter and somewhat cleaner.  The $\eta=0$ line
still appears to be a good boundary to separate stars that are most
likely in the halo from those in the disk or thick disk.  However,
the boundary for secure halo stars can be placed at $\eta=0.5$,
rather than at $\eta=1.0$, as was adopted by \citet{gould03halo}.

\section{Halo Parameters
\label{sec:halo}}

We fit the halo stars in rNLTT (with the newly recalibrated photometry)
to the 28-parameter halo model of \citet{gould03halo}.  We apply almost
exactly the same procedures as in the original work, which we review
very briefly here.

\subsection{Review of Model and Method
\label{sec:review}}

The model contains 13 parameters to describe the luminosity function (LF),
$\Phi(M_{V,i}),\ i= 3\ldots 15$,
one for each 1-magnitude bin from $M_V=3$ to $M_V=15$.  It has 9 parameters
describing the halo velocity ellipsoid including 3 for bulk motion relative
to the Sun $(U_i)$ and 6 for the velocity dispersion tensor [3 dispersions 
$c_{ii}$ and 3 correlation coefficients, 
$\tilde c_{ij}\equiv c_{ij}/(c_{ii}c_{jj})^{1/2}$]. 
The halo color-magnitude relation (CMR) is
described by 2 parameters $(a,b)$, i.e., $M_V=a(V-J)_0 + b$.  There are
2 parameters describing the halo density profile $(\nu,\kappa)$, i.e.,
$\rho = \rho_0(R/R_0)^{-\nu}\exp(-\kappa|z|)$.  Here, $R$ is Galactocentric
distance, $R_0=R({\rm sun})$, $z$ is distance from the plane, $\nu$ is
the halo density power law, and $\kappa$ is the inverse scale height.
Finally, there are 2 parameters describing the sample completeness,
$V_{\rm break}$ and $f_{\rm break}$.  The sample is assumed 100\%
complete for $V<12$ and then to fall linearly to $f_{\rm break}$ at
$V_{\rm break}$, and then to fall linearly from there to zero at 
$V=20$.  Of these 28 parameters, one is held fixed at $U_2=-216.6\,\kms$
in order to enforce an otherwise unconstrained distance scale.

The best-fit model is found by maximizing the likelihood
\begin{equation}
\ln L = \sum_{k=1}^{N_{\rm det}} \ln \{P_k[z^m(z^m_{\rm obs})]{\cal J}\}
-N_{\rm exp},
\label{eqn:liklihood}
\end{equation}
where $N_{\rm det}$ is the number of stars in the sample, $N_{\rm exp}$
is the number expected in the model, $z^m$ are the $m=6$ phase space
coordinates per star in the model, $z^m_{\rm obs}$ are the $m=6$
observables $(l,b,\mu_l,\mu_b,V,J)$ from which these coordinates are
inferred, $P_k$ is the probability that the $k$ star will have
the phase-space coordinates that have been inferred from the observations
given the model parameters, and ${\cal J}$ is the Jacobian of the
transformation from the observables to the phase-space coordinates.
(Note that the rows in the second matrix of \citealt{gould03halo}'s
eq.~[6] should be reversed, but that the final result is correct.)
The number expected $(N_{\rm exp})$ in each trial model is evaluated
by Monte Carlo integration, which is populated at 
100 times higher density than the actual sample to suppress Poisson
errors.  However, to avoid fluctuations when comparing one model to
another, the random positions and velocities are chosen only
once, and then are assigned different weights depending on the
model parameters.

\citet{gould04} found a bug in the likelihood code used by \citet{gould03halo}
and gave revised parameters corrected for this bug.  However, most of the
corrections were very small.

\subsection{Changes in Approach
\label{sec:changes}}

	We implement two major changes relative to \citet{gould03halo}'s
treatment (in addition to using the recalibrated photometry).
First, based on inspection of Figure \ref{fig:rpm}, we
select stars in the interval $0.5\leq \eta \leq 4.15$.
This yields a sample of 5042 ``secure halo stars'', somewhat larger
than the 4588 analyzed by \citet{gould03halo} (and corrected to
4564 in \citealt{gould04}).

	Second, we discovered that \citet{gould03halo}'s likelihood
maximization routine was too ``stiff'' to probe the effect of
simultaneous changes in the two color-magnitude parameters, $(a,b)$,
and the LF, $\Phi(M_{V,i})$.  
When either or both of the color-magnitude parameters change, this
affects all of the LF
parameters.  Unless all can be varied simultaneously in just 
the right way,
the true maximum cannot be
located.  We address this problem using the method of ``hybrid
statistical errors'' of \citet{an02}.  We hold the two color-magnitude
parameters $A_i^{(a,b)}=(a,b)$ fixed at a grid of values and evaluate
the likelihood $L$ 
for the remaining 25 free parameters $A_k^{\rm (remain)}$.
We then find the inverse covariance matrix for the color magnitude
parameters,
\begin{equation}
B^{(a,b)}_{ij} = -{\partial^2 L\over \partial 
A_i^{(a,b)}\partial A_j^{(a,b)}}
\label{eqn:bijab}
\end{equation}
and so obtain the restricted covariance matrix $C^{(a,b)}=[B^{(a,b)}]^{-1}$.
We also find the gradient of all 27 parameters with respect to the 
two color-magnitude parameters over the grid of solutions,
$\partial A_k/\partial A_i^{(a,b)}$.  We determine the
covariance matrix, $C_{ij}^{\rm (remain)}$ 
of the remaining 25 free parameters (with $[a,b]$ held
fixed at their likelihood maximum) using the bootstrap technique.
Finally, we find the hybrid covariance matrix
\begin{equation}
C_{ij} = C_{ij}^{\rm (remain)} + \sum_{m,n} C_{mn}^{(a,b)} 
{\partial A_i\over \partial A_m^{(a,b)}}
{\partial A_j\over \partial A_n^{(a,b)}}.
\label{eqn:hybrid}
\end{equation}

\subsection{Results
\label{sec:results}}

As discussed below equation (\ref{eqn:Vcor}), the
primary changes in the data set are to make the faint stars fainter
absolutely, but to make the tracks of subdwarf (and main-sequence)
stars {\it brighter} at fixed color.  Hence, we expect that the 
main changes in the fit
will be to make the CMR fainter and shallower,
and to move the peak in the LF toward fainter magnitudes.  Table 1
shows that these are indeed the main effects.  The first two columns give
the parameter name and units.  The next two give the best fit values
as determined using the old and new photometric calibration, respectively.
The final two columns give the respective errors.  

Figure \ref{fig:lf} shows the new LF together with several other
determinations from the literature.  Comparing this figure to
Figure 2 from \citet{gould04} (of which it is an updated version),
one sees that the rNLTT-based LF has moved toward very good agreement
with the LF of \citet{dahn} (DLHG) as renormalized by \citet{gould03halo}
using the results of \citet{crb} (CRB).  Indeed, with the
exception of the final DLHG/CRB point, the two are in agreement at the
$1\,\sigma$ level.  This resolves an important puzzle: the two determinations
are both essentially local and so should be similar. It would be quite
surprising if their peaks were separated by a magnitude, as appeared to
be the case before the rNLTT photometry was recalibrated.

There are a few other changes that should be noted as well.  First,
while most of the error bars are similar with the old and new photometry,
those of the two color-magnitude parameters $(a,b)$ and of the three
velocity dispersions $(c_{ii})$ have all grown significantly.  As mentioned
above, the previous algorithm was too stiff to properly evaluate the
errors in $(a,b)$, so it is not surprising that they have grown.  This
growth is also responsible for the increase in the $c_{ii}$ errors through
the second term in equation (\ref{eqn:hybrid}), which was not previously
incorporated.  Of course, this term also increases the errors of all other
parameters, but it turns out that these other
increases are mostly not significant
relative to the errors given by the first term.

Another significant change is that the break in the completeness function,
$V_{\rm break}$ has moved about 1/2 mag fainter.  This is also not
surprising given that the whole photometric calibration has moved
fainter at the faint end.  However, the completeness level at this
break has also moved lower, and this has disturbing consequences,
as we discuss in the next section.

Finally, we note that the two bulk velocity parameters and three
velocity-ellipsoid correlation coefficients
$(U_1,U_3,\tilde c_{12},\tilde c_{13},\tilde c_{23})
=(10.6\pm 1.4,-6.4\pm 1.8,0.017\pm 0.015,-0.010\pm 0.017,-0.036\pm 0.025)$
remain very close to the values expected in an axisymmetric Galaxy,
$(10.0,-7.2,0,0,0)$, when account has been taken of the Sun's motion
with respect to the local standard of rest 
$(-10.00\pm 0.36,+7.17\pm 0.38)\,\kms$ in the radially outward and vertical
directions \citep{dehnen98}.  This yields a $\chi^2=4.04$ for 
5 degrees of freedom, almost identical to the value 3.97 obtained by
\citet{gould03granularity} from the halo solution of \citet{gould03halo}.
This implies that the constraints on the granularity of the stellar
halo derived by \citet{gould03granularity} from this $\chi^2$ determination
remain unaltered.

\section{Two Puzzles
\label{sec:puzzles}}

While it is comforting to see the old puzzle regarding the peak of the halo
LF resolved (see \S~\ref{sec:results}), the halo solution derived using
recalibrated rNLTT photometry presents two new puzzles.  These concern
conflicts with independent determinations of NLTT completeness and
of the $(V,V-J)$ CMR.

\subsection{Completeness
\label{sec:completeness}}

The completeness fraction ($f_{\rm break} = 0.38$)
at the completeness break point ($V_{\rm break}=18.8$) seems quite low.
\citet{gould03halo} had argued that a somewhat higher value
was consistent with what was then otherwise known about the completeness
of NLTT (see also \citealt{gould04}).
 However, not only is the new value of $f_{\rm break}$ lower,
but \citet{lepine05} have shown, using their own independent and very
complete northern-sky proper-motion catalog, that 
at high latitude, $b>15^\circ$, NLTT is 85\% complete
at $V=18.5$.
We now explore several ideas to resolve this conflict, but the 
executive summary is: none are successful.

First, the \citet{lepine05} completeness estimate strictly applies
only for proper motions $\mu\geq 250\,\masyr$, whereas rNLTT goes
down to $180\,\masyr$.   Completeness falls to 79\% at $V=18.5$
for $\mu\geq 200\,\masyr$.  However, this ``incompleteness'' 
is simply due to NLTT's $20\,\masyr$ errors combined
with its $180\,\masyr$ threshold: stars that are mismeasured below
this threshold due to normal statistical errors are not included in the
catalog.  This effect is already taken into account in our likelihood
procedure.  Moreover, even if it were not accounted for, the effect
is much smaller than the discrepancy and so could not account for
it in any case.

Second, rNLTT is less complete than NLTT, and the halo sample used here
is less complete than rNLTT, primarily 
because stars without $J$-band data are excluded.
However, as shown by \citet{salim03}, rNLTT is about 97\% complete
relative to NLTT down to $R_{\rm NLTT}= 17$ and 95\% complete
at $R_{\rm NLTT}=18$  These magnitudes correspond roughly to
$V\sim 17.5$ and $V\sim 18.5$, respectively.  It is true that 
the completeness
at the faint end is substantially worse if one excludes stars
without $J$-band data.  However, as was argued by \citet{salim03}
and strikingly confirmed by \citet{lepine05}, almost all faint NLTT
stars that lack 2MASS counterparts are white dwarfs, not subdwarfs.
Hence, incompleteness of rNLTT relative to NLTT can explain at most
a few percent of the effect.

Third, the form of the completeness function adopted by \citet{gould03halo},
and summarized above in \S~\ref{sec:review}, could in principle be too 
simple to capture the evolution of NLTT completeness over 8 magnitudes.  In 
fact, however, from Figure 22 of \citet{lepine05}, this form actually looks
quite appropriate, except that the parameters should rather be
$f_{\rm break} = 0.8$, $V_{\rm break}=18.7$.  Nevertheless, we further
test this possibility by eliminating the faintest stars in the likelihood 
sample $V>18.5$, and refitting with a simple 1-parameter completeness
function, $f_{\rm break}$ at $V=18.5$.  We find a best fit
$f_{\rm break}=0.39$.  If we enforce $f_{\rm break}=0.80$ (to take account
of both the 15\% NLTT incompleteness and the 5\% incompleteness of rNLTT
relative to NLTT at the faint end), then the likelihood
 falls by 25, which means that this potential solution is ruled
out at the $7\,\sigma$ level.  To be conservative, we repeat this
exercise with the cutoff at $V=18$, but still find that the best-fit
$f_{\rm break}=0.41$ is preferred at the $7\,\sigma$ level over the
independently-determined value of $f_{\rm break}=0.85$.

Is it possible that \citet{lepine05} overestimated the completeness of
NLTT?  We believe not.  At high latitudes, they failed to detect only
1\% of NLTT stars.  They tracked down the reason for this failure
in each case and found in essentially all cases that the star was
contaminated by a random field star
in their own (circa 1990) second epoch, but was free
from contamination in NLTT's (circa 1970) second epoch.  They inferred
that they were missing an additional 1\% from stars corrupted in
the common NLTT/LSPM (circa 1950) first epoch.  Hence, LSPM is itself
nearly 100\% complete and so forms an excellent template against
which to measure the completeness of NLTT.

If $f_{\rm break}$ is forced to high values, then most of the parameters
in the halo fit remain unchanged.  However, the LF is suppressed by
an amount that decreases from a factor 1 at the bright end to a factor
$\sim (f_{\rm break}/0.8)$ at the faint end. Such an adjustment would engender
a conflict between the DLHG/CRB LF and the one derived from the halo 
fit.  However, as we discuss in the next section, this in itself is
not a strong argument against making the adjustment.  The principal
argument is simply that it leads to a poor fit to the rNLTT data.
In brief, at present we see no clear path to resolving this conflict.

\subsection{Color-Magnitude Relation (CMR)
\label{sec:cmr}}

The second puzzle concerns the CMR. 
\citet{gould03halo} derived a CMR from stars with trigonometric
parallaxes taken from \citet{monet92} and \citet{gizis97} that
was quite consistent with the CMR derived from the halo likelihood
fit.  Indeed, the two CMRs lie almost exactly on top of one another
in his Figure 3.  However, after photometric recalibration, our
new CMR has changed, while the parallax-based CMR should remain
essentially the same.  In fact, we slightly change our selection
of parallax stars to be consistent with our ``secure halo''
criterion, $0.5\leq \eta \leq 4.15$, but the impact of this
change is expected to be small.

	Before restricting attention to the halo stars, we first
find the mean photometric offset between the CCD $V$ magnitudes
given by \citet{monet92} and \citet{gizis97} and
the recalibrated rNLTT $V$ magnitudes for the entire sample 
of 58 rNLTT-parallax
stars with $V>12$.  We find a mean difference $0.03\pm 0.04$ mag,
in the sense that recalibrated rNLTT is brighter.  The fact
that the two are consistent at the $1\,\sigma$ level serves as
a sanity check on the SDSS-based calibration, although the
smaller number of parallax stars makes the uncertainty in this
test uncompetitive with SDSS.

	As \citet{gould03halo} did, we eliminate the reddest
star on the ground that the CMR for the latest-type stars 
may deviate from a straight line and take a turn toward the red.
(There are too few very late subdwarfs in the rNLTT sample to
test this conjecture, so it is better to just eliminate this
star from the comparison.)  We use the CCD photometry both
to make the selection (i.e., determine $\eta$), and to estimate
the absolute magnitude.  This yields a sample of 23 rNLTT-parallax 
halo stars.
We find that we must add an error $\sigma(M_V)=0.72$ in quadrature to
the errors propagated from the given parallax errors, in order
to achieve $\chi^2/{\rm dof}=1$.  Fitting to the form $M_V=a(V-J)_0 + b$,
we find,
\begin{equation}
a = 3.668\pm 0.396,\qquad b=0.324\pm 1.201,\qquad \rho=-0.9906
\qquad ({\rm parallax\ CMR})
\label{eqn:parallaxcmr}
\end{equation}
compared to the values
\begin{equation}
a = 3.339\pm 0.027,\qquad b=0.921\pm 0.073,\qquad \rho=-0.8948
\qquad ({\rm halo\ fit})
\label{eqn:rnlttcmr}
\end{equation}
obtained in the halo fit in \S~\ref{sec:results}.  Here $\rho$
is the correlation coefficient.  Figure \ref{fig:cmd} shows the
parallax data with various CMRs.

\begin{figure}
\epsscale{0.8}
\plotone{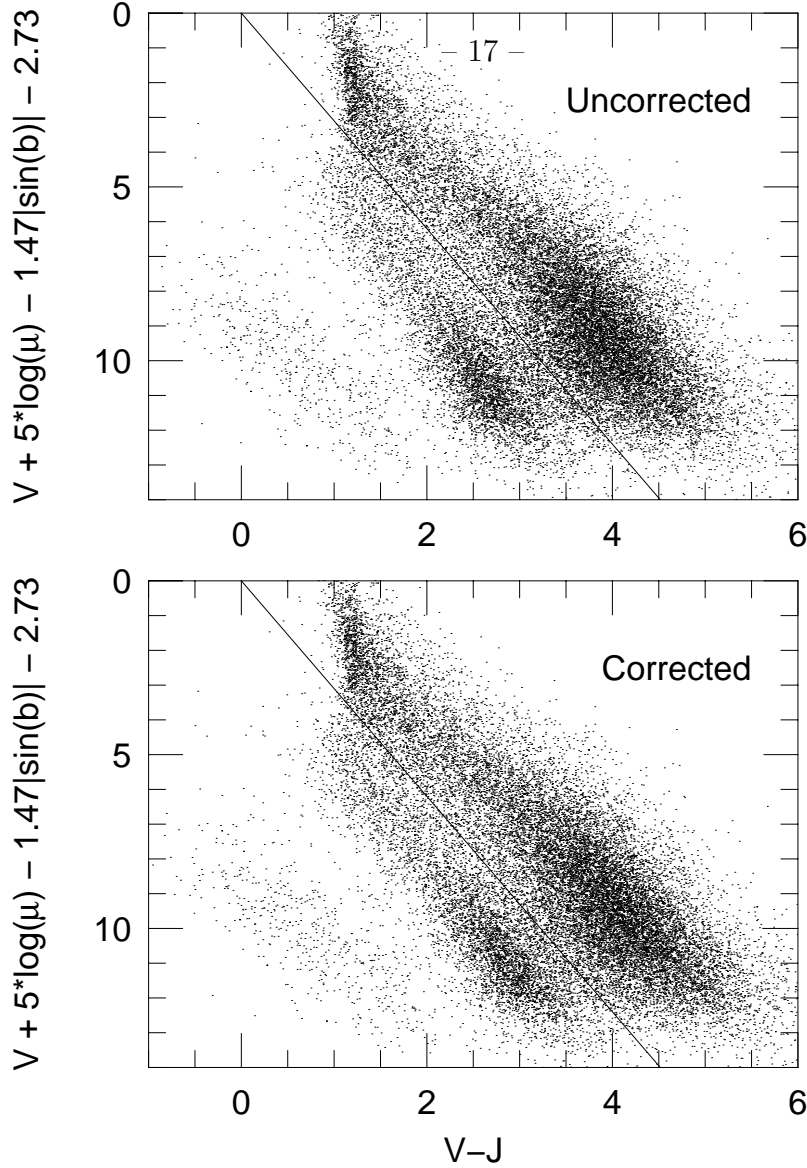}
\caption{\label{fig:rpm}
Reduced proper motion (RPM) diagram before (upper) and after (lower)
calibration of rNLTT $V$ band photometry.   Following \citet{salim03},
the RPM includes a term that depends on Galactic latitude ($b$) as
well as proper motion ($\mu$), 
$V_{\rm RPM,adj} \equiv V + 5\log(\mu) -1.47|\sin(b)| - 2.73$.  This quantity 
enters the discriminant $\eta \equiv V_{\rm RPM,adj} - 3.1(V-J)$. 
\citet{salim03}
adopted $\eta=0$ ({\it solid line}) as a boundary between halo stars
(below) and disk and thick-disk stars (above).  The new calibration
moves the faint end of both the halo and disk sequences 
toward the red and straightens
and somewhat cleans up the ``trough'' between the halo and disk
populations.  We still regard $\eta=0$ as a good boundary between
disk and halo stars.
}\end{figure}

\begin{figure}
\plotone{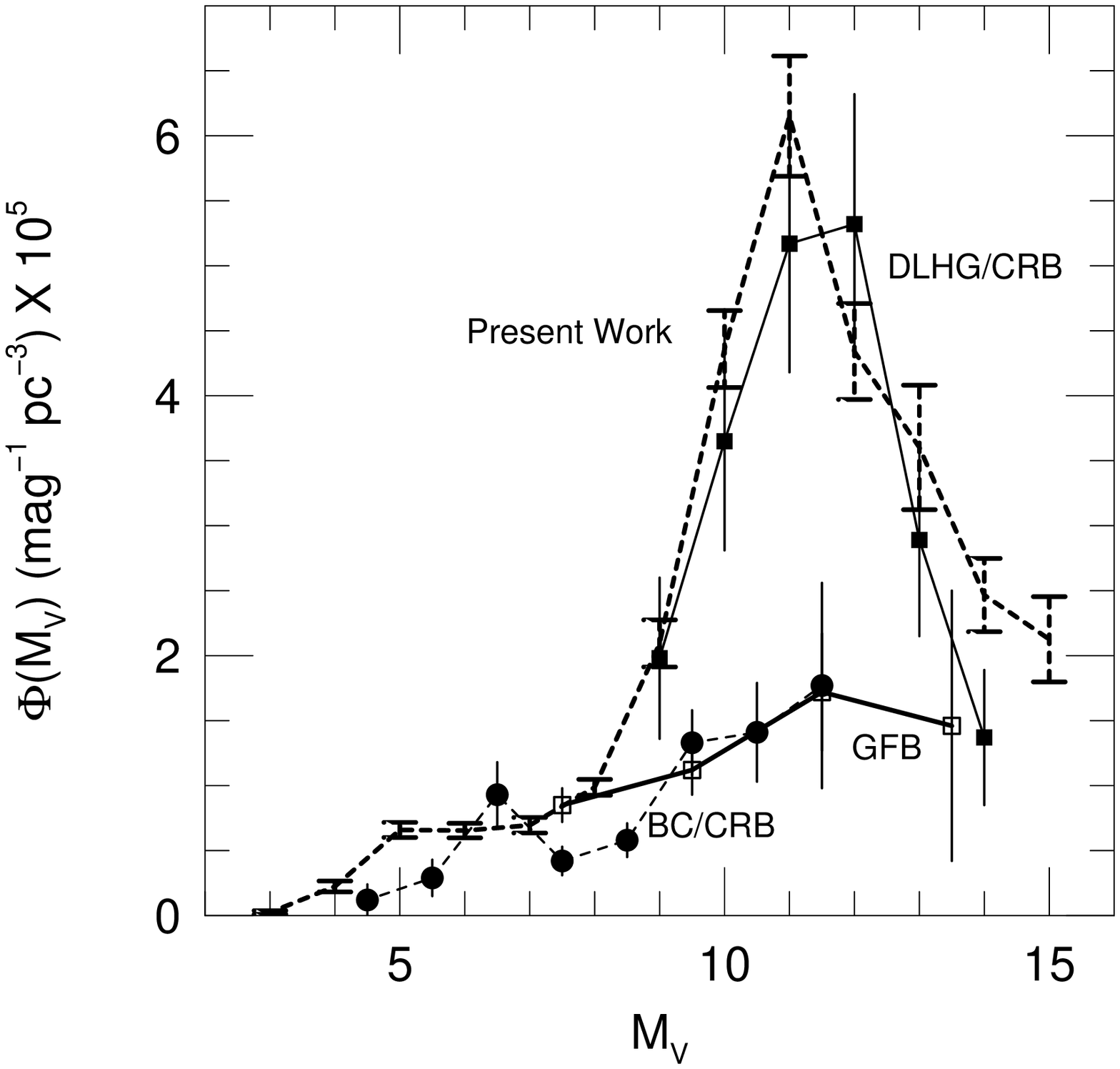}
\caption{\label{fig:lf}
Comparison of the halo luminosity function (LF) derived in the present
work ({\it bold dashed curve}) using calibrated rNLTT photometry
with several previous determinations.  The rNLTT LF is now in good
agreement with the parallax-based determination 
of \citet{dahn} as renormalized by \citet{gould03halo}
based on results from \citet{crb}, which is shown as a {\it solid curve}
(DLHG/CRB).  These are both local measurements, so they should agree,
but in an analysis prior to the new calibration \citep{gould03halo,gould04}, 
the rNLTT
determination peaked about 1 mag brighter than DLHG/CRB. 
The measurement of \citet{gould97} (GFB) agrees with
the present work at bright magnitudes but shows a much weaker
peak.  However, it is based on a distant sample, so in principle may
be probing a different population.  Also shown is the determination
of \citet{bc} as renormalized by \citet{gould03halo} using CRB
(BC/CRB).
}\end{figure}

\begin{figure}
\plotone{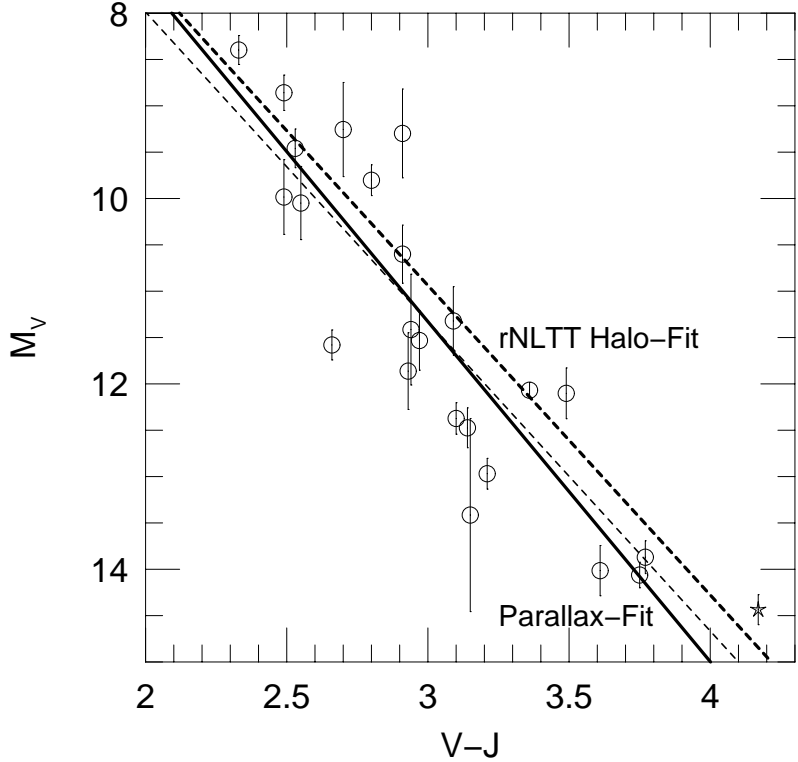}
\caption{\label{fig:cmd}
Parallax-based absolute magnitudes $M_V$ and $(V-J)$ colors
of 23 stars ({\it circles}) used to estimate the color-magnitude
relation (CMR), given by eq.\ (\ref{eqn:parallaxcmr}) and 
shown as a bold solid curve.  Also
shown is one star that is excluded from the fit ({\it star}).  
The error bars on the points reflect only the parallax errors and
not the 0.72 mag ``cosmic scatter'' that was added in the fitting
process.  The
bold dashed curve shows the CMR derived from the fit to rNLTT halo
stars presented in this paper.  When the parallax-fit is forced
to have the same slope as the rNLTT halo model (which has much
higher statistical weight), its track is given by the thin dashed curve.
When the parallax data are combined with the rNLTT halo fit, the resulting
CMR is parallel to and roughly halfway between the two dashed curves.
Its form is given by eq.~(\ref{eqn:cmrnet}), but it is not shown in this
figure to avoid clutter.
}\end{figure}

	While the slopes and zero points of equations 
(\ref{eqn:parallaxcmr}) and (\ref{eqn:rnlttcmr}) are each consistent
at the $1\,\sigma$ level, the two are highly correlated, so that
the relations as a whole are mildly inconsistent, with 
$\Delta\chi^2=4.0$.  To interpret this discrepancy, we must first
ask how the slope and zero point of the halo-fit CMR depend on
the data and the  assumptions.  The slope is determined fairly
directly from the rNLTT data themselves and primarily reflects the
slope of the ``subdwarf sequence'' seen in the RPM diagram
(Fig.\ \ref{fig:rpm}).  On the other hand, the zero point is
directly determined by fixing $U_2=-216.6\,\kms$.  If this
value had been fixed 10\% faster, then all of the 5 other velocity
parameters would have increased in lock step by the same 10\%,
and all the inferred distances would have increased by the same
amount.  This rigid scaling occurs because (apart from a very
slight effect in the extinction prescription) the only
way that the distance scale enters the likelihood calculation
is through $U_2$.  This distance scaling then implies that 
the LF would have been reduced uniformly by a factor
$(1.1)^{-3}$.  However, the value of $U_2=-216.6\,\kms$ was
adopted directly from the statistical-parallax solution of
\citet{gp} and is therefore to some extent arbitrary.  First,
that measurement had a $1\,\sigma$ statistical error of $12.5\,\kms$.
Second, as \citet{gp} (and references therein) note, 
different selection criteria will yield halo samples with values of
$U_2$ that differ by of order $10\,\kms$.  \citet{gould03halo} argued
that the ``$\eta$'' selection criterion used in his (and our) halo
analysis was most similar to that of \citet{gp}, but there could
still be some differences between the samples.  Combining these two effects,
the true value of $U_2$ might plausibly be different from the 
adopted one by of order $15\,\kms$, corresponding to 0.15 mag in the
CMR zero point.

	Hence, when combining the two CMRs, we should insist
on a common slope, but initially allow the two zero points to
differ.  Since the halo-fit's slope error is about 15 times smaller
than that of the parallax determination,
this amounts to fixing the parallax-based slope at the halo-fit
value.  The parallax-based zero point then becomes 
$b=1.31\pm 0.16$, which is 0.29 mag fainter than the
halo-fit $b=0.92$.  The direction of this discrepancy may seem
surprising at first sight because the photometric recalibration
moved the halo stars toward {\it fainter} mags, yet the CMR based on
these stars has now become {\it brighter} than the parallax-based CMR.
The reason is that the halo stars became not only {\it fainter}
but {\it redder}, and because the slope of the CMR is larger than
unity, the latter is the larger effect.  Hence, at {\it fixed color}
(which is what is important for the CMR), the halo stars have become
{\it brighter}, even though they are fainter absolutely.

	Having reasoned through the problem heuristically and verified
that the two CMRs give qualitatively similar results, we combine them
in a formally rigorous way by averaging the two $(a,b)$ vectors,
each weighted by its inverse covariance matrix.  To take account
of the external error due to uncertainty in the \citet{gp} velocity
input, we first add 0.15 mag to the halo-fit uncertainty in $b$.
We then find
\begin{equation}
a=3.3438\pm 0.0269\qquad  b = 1.091\pm 0.134,\qquad \rho=-0.543
\qquad \rm (combined\ determinations) 
\label{eqn:cmrnet}
\end{equation}

	At the ``center of mass'' of the halo-star color distribution,
$V-J=2.4$, this relation is $0.17\pm 0.10$ fainter than the
CMR derived from the halo fit alone (but assuming a known $U_2$).
Hence, taking account of this calibration (and the added uncertainty
in the $U_2$ constraint) the velocities should all be smaller
by a factor $0.93\pm 0.04$, while the LF should be higher
by a factor $1.26\pm 0.16$.  These changes are not large, but they
do mean that one cannot demand too close agreement between the
LF derived from the rNLTT sample and that derived from local
star counts by DLHG/CRB (see \S~\ref{sec:completeness}).

\acknowledgments

Work by AG and JAK was supported by grant AST 02-01266.  Work by
JC was supported by an Ohio State University Presidential Fellowship.


\clearpage

\begin{deluxetable}{l l r r r r}
\tablecaption{Halo Parameters with Calibrated rNLTT Photometry\label{tab:haloparm}}
\tablewidth{0pt}
\tablehead{

\colhead{Parameter} &
\colhead{Units} &
\colhead{Old} &
\colhead{New} &
\colhead{Old} &
\colhead{New} \\
& &
\colhead{Value} &
\colhead{Value} &
\colhead{Error} &
\colhead{Error} }
\startdata
$\Phi(M_V=3)$ & $10^{-5}{\rm pc^{-3}}$ &      $   0.030 $ & $    0.022 $ & $  0.010 $ & $  0.014$ \\
$\Phi(M_V=4)$ & $10^{-5}{\rm pc^{-3}}$ &      $   0.170 $ & $    0.244 $ & $  0.030 $ & $  0.042$ \\
$\Phi(M_V=5)$ & $10^{-5}{\rm pc^{-3}}$ &      $   0.490 $ & $    0.647 $ & $  0.060 $ & $  0.058$ \\
$\Phi(M_V=6)$ & $10^{-5}{\rm pc^{-3}}$ &      $   0.610 $ & $    0.654 $ & $  0.060 $ & $  0.056$ \\
$\Phi(M_V=7)$ & $10^{-5}{\rm pc^{-3}}$ &      $   0.640 $ & $    0.675 $ & $  0.060 $ & $  0.060$ \\
$\Phi(M_V=8)$ & $10^{-5}{\rm pc^{-3}}$ &      $   0.860 $ & $    0.987 $ & $  0.090 $ & $  0.061$ \\
$\Phi(M_V=9)$ & $10^{-5}{\rm pc^{-3}}$ &      $   2.240 $ & $    2.126 $ & $  0.170 $ & $  0.181$ \\
$\Phi(M_V=10)$ & $10^{-5}{\rm pc^{-3}}$ &     $   4.660 $ & $    4.314 $ & $  0.420 $ & $  0.296$ \\
$\Phi(M_V=11)$ & $10^{-5}{\rm pc^{-3}}$ &     $   4.500 $ & $    6.208 $ & $  0.420 $ & $  0.463$ \\
$\Phi(M_V=12)$ & $10^{-5}{\rm pc^{-3}}$ &     $   2.570 $ & $    4.190 $ & $  0.320 $ & $  0.369$ \\
$\Phi(M_V=13)$ & $10^{-5}{\rm pc^{-3}}$ &     $   2.070 $ & $    3.727 $ & $  0.370 $ & $  0.479$ \\
$\Phi(M_V=14)$ & $10^{-5}{\rm pc^{-3}}$ &     $   1.660 $ & $    2.376 $ & $  0.330 $ & $  0.282$ \\
$\Phi(M_V=15)$ & $10^{-5}{\rm pc^{-3}}$ &     $   1.530 $ & $    2.126 $ & $  0.450 $ & $  0.328$ \\
$U_1$ & ${\rm km\,s^{-1}}$ &                 $   8.500 $ & $   10.582 $ & $  2.200 $ & $  1.408$ \\
$U_3$ & ${\rm km\,s^{-1}}$ &                 $  -7.500 $ & $   -6.395 $ & $  2.400 $ & $  1.787$ \\
$c_{11}^{1/2}$ & ${\rm km\,s^{-1}}$ &        $ 167.900 $ & $  165.580 $ & $  1.400 $ & $  2.426$ \\
$c_{22}^{1/2}$ & ${\rm km\,s^{-1}}$ &        $ 113.000 $ & $  115.279 $ & $  1.700 $ & $  3.481$ \\
$c_{33}^{1/2}$ & ${\rm km\,s^{-1}}$ &        $  88.600 $ & $   89.342 $ & $  1.900 $ & $  2.259$ \\
$\tilde c_{12}$ & &                          $   0.008 $ & $    0.017 $ & $  0.014 $ & $  0.015$ \\
$\tilde c_{13}$ & &                          $   0.014 $ & $   -0.010 $ & $  0.023 $ & $  0.017$ \\
$\tilde c_{23}$ & &                          $  -0.039 $ & $   -0.036 $ & $  0.026 $ & $  0.025$ \\
$a$ & &                                      $   3.590 $ & $    3.339 $ & $  0.010 $ & $  0.027$ \\
$b$ & &                                      $   0.690 $ & $    0.921 $ & $  0.010 $ & $  0.073$ \\
$f_{\rm break}$ & &                          $   0.500 $ & $    0.384 $ & $  0.060 $ & $  0.027$ \\
$V_{\rm break}$ & &                          $  18.270 $ & $   18.809 $ & $  0.040 $ & $  0.102$ \\
$\nu$ & ${\rm kpc^{-1}}$ &                   $   2.700 $ & $    3.257 $ & $  1.000 $ & $  0.734$ \\
$\kappa$ & ${\rm kpc^{-1}}$ &                $   0.019 $ & $    0.059 $ & $  0.057 $ & $  0.054$ 
\enddata
\end{deluxetable}


\clearpage

\begin{thebibliography}{}

\bibitem[Abazajian et al.(2005)]{abaz05} Abazajian, K. 2005, \aj,
submitted (astroph/0410239)

 
\bibitem[An et al.(2002)]{an02} An, J.H., et al. 2002, \apj, 572, 521

\bibitem[Bahcall \& Casertano(1986)]{bc} 
Bahcall, J.N., \& Casertano, S.\ 1986, \apj, 308, 347

\bibitem[Casertano et al.(1990)]{crb} Casertano, S.,
Ratnatunga, K., \& Bahcall, J.N.\ 1990, \apj, 357, 435

\bibitem[Chaname \& Gould(2004)]{chaname04}
Chaname, J., \& Gould, A.\ 2004, \apj, 601, 289

\bibitem[Dahn et al.(1995)]{dahn} Dahn, C.C., Liebert, J.W., Harris, H., 
\& Guetter, H.C.\ 1995, 
p.\ 239, An ESO Workshop on: the Bottom of the Main Sequence and Beyond,
C.G.\ Tinney ed.\ (Heidelberg: Springer)

\bibitem[Dehnen \& Binney(1998)]{dehnen98} Dehnen, W. \&  Binney, J.J. 
1998, \mnras, 298, 387


\bibitem[ESA(1997)]{esa97} European Space Agency (ESA). 1997, The Hipparcos and
Tycho Catalogues (SP-1200; Noordwijk: ESA)

\bibitem[Gizis(1997)]{gizis97} Gizis, J.~E. 1997, \aj, 113,508



\bibitem[Gould(2003a)]{gould03halo}
Gould, A.\ 2003a \apj, 583, 765

\bibitem[Gould(2003b)]{gould03granularity}
Gould, A.\ 2003b, \apjl, 592

\bibitem[Gould(2004)]{gould04}
Gould, A.\ 2004 \apj, 607, 653




\bibitem[Gould et al.(1997)]{gould97} Gould, A., Flynn, C., \& Bahcall, J.N.
1997, \apj, 503, 798


\bibitem[Gould \& Popowski(1998)]{gp} Gould, A. \& Popowski, P. (1998)
\apj, 508,844

\bibitem[Gould \& Salim(2003)]{gould03bright}
Gould, A., \& Salim, S., 2003, ApJ, 582, 1001




\bibitem[H{\o}g et al.(2000)]{t2} H{\o}g, E.~et al.\ 2000, \aap, 355, L27




\bibitem[Juri\'c et al.(2002)]{juric02} Juri\'c, M. et al. 2002, \aj, 124, 1776

\bibitem[Liebert et al.(1988)]{ldm88} Liebert, J., Dahn, C.C., \& Monet, D.G.
1988, \apj, 332, 891

\bibitem[Kawka et al.(2004)]{kawka04} Kawka, A., Vennes, S., Thorstensen, J.R.
2004, \aj, 127 1702

\bibitem[L\'{e}pine \& Shara(2005)]{lepine05} 
L\'{e}pine, S. \& Shara, M.M. 2005, \aj, in press (astroph/0412070)




\bibitem[Lowrance et al.(2003)]{lowrance03}
Lowrance, P.J., Kirkpatrick, J.D., Reid, I.N., Kelle, L.C., \& Liebert, J.
2003, \apj, 584, L98

\bibitem[Luyten(1979, 1980)]{luyten7980} Luyten, W.\ J.\ 1979, 1980, New Luyten
Catalogue of Stars with Proper Motions Larger than Two Tenths of an Arcsecond
(Minneapolis: University of Minnesota Press)




\bibitem[Monet(1998)]{usnoa2} Monet, D.~G.\ 1998, American Astronomical
Society Meeting, 193, 112003

\bibitem[Monet et al.(1992)]{monet92}
Monet, D.G., Dahn, C.C., Vrba, F.J., Harris, H.C., Pier, J.R.,
Luginbuhl, C.B., \& Ables, H.D. 1992, \aj, 103, 638

\bibitem[Monet et al.(2003)]{monet03}
Monet et al. 2003, AJ, 125, 984


\bibitem[Reid et al.(2004)]{reid04} Reid, I.N. 2004, \aj, 128, 463







\bibitem[Salim \& Gould(2002)]{salim02} 
Salim, S. \& Gould, A., 2002, \apj, 575, L83

\bibitem[Salim \& Gould(2003)]{salim03}
Salim, S., \& Gould, A.\ 2003 \apj, 582, 1011



\bibitem[Skrutskie et al.(1997)]{2mass} Skrutskie, M.~F.~et al.\ 1997, in The
Impact of Large-Scale Near-IR Sky Survey, ed. F. Garzon et al (Kluwer:
Dordrecht), p.\ 187


\bibitem[Vennes \& Kawka(2003)]{vennes03}Vennes, S. \& Kawka, A. 2003,
\apj, 586, L95

\bibitem[Weis(1996)]{weis96} Weis, E.W. 1996, \aj, 112, 2300

 

\bibitem[Yong \& Lambert(2003)]{yong03} 
Yong, D. \& Lambert, D.L. 2003, \aj, 126 2449




\end{thebibliography}
\end{document}